\documentclass[doublecol,article,superscriptaddress,floatfix]{epl2}
\usepackage{float,graphicx,bm,dcolumn,caption,subcaption,epsf,amsmath,amssymb,wasysym,verbatim,color,enumitem,ragged2e}
\usepackage[numbers]{natbib}

\title{Coevolving nonlinear voter model with triadic closure}
\shorttitle{Coevolving nonlinear voter model with triadic closure}

\author{Tomasz Raducha\inst{1,2}\thanks{\email{tomasz.raducha@fuw.edu.pl}}
	\and Byungjoon Min\inst{2,3}
	\and Maxi San Miguel\inst{2}\thanks{\email{maxi@ifisc.uib-csic.es}} }
\shortauthor{T. Raducha \etal}

\institute{
	\inst{1} Institute of Experimental Physics, Faculty of Physics, University of Warsaw,
		Pasteura 5, 02-093 Warsaw, Poland \\
	\inst{2} IFISC, Instituto de F\'isica Interdisciplinar y Sistemas Complejos (CSIC-UIB),
		Campus Universitat Illes Balears, E-07122 Palma de Mallorca, Spain \\
	\inst{3} Department of Physics, Chungbuk National University,
		Cheongju, Chungbuk, 28644 South Korea
}

\date{\today}
\abstract{We study a nonlinear coevolving voter model with triadic closure local rewiring. 
We find three phases with different topological properties and configuration in the 
steady state: absorbing consensus phase with a single component, absorbing fragmented phase with
two components in opposite consensus states, and a dynamically active shattered phase
with many isolated nodes. This shattered phase, which does not exist for a coevolving 
model with global rewiring, has a lifetime that scale exponentially with system size. 
We characterize the transitions between these phases in terms of the size of the 
largest cluster, the number of clusters, and the magnetization. Our analysis 
provides a possible solution to reproduce isolated parts in adaptive networks 
and high clustering widely observed in social systems.
}

\begin{document}
\maketitle



\section{Introduction}\label{section:int}
Network dynamics \cite{dorogovtsev2002evolution,boccaletti2006complex,barrat2008dynamical,albert2002statistical,castellano2009statistical} often refers to the dynamics \emph{of} the network creation or evolution, but also to the dynamics \emph{on} the network, where the dynamics of the states of the nodes is affected by the topological structure of a fixed underlying network. However, the structure of the network among interacting agents evolves dynamically in response to the state of the nodes. Therefore, dynamics \emph{of} and \emph{on} the network are dynamically coupled processes occurring in comparable time scales. The coupling of these processes has been referred as Coevolution of node states and network structure \cite{zimmermann2001cooperation,zimmermann2004coevolution}.

In order to understand this coevolutionary dynamics
of complex systems, there have been several studies which incorporate
both the change of networks and the dynamical processes on networks
\cite{vazquez2008generic,gross2006epidemic,holme2006nonequilibrium,gross2008adaptive,biely2009socio}.
Coevolutionary dynamics has been applied in a variety of different fields
ranging form classical spin models \cite{biely2009socio,toruniewska2016unstable},
opinion formation \cite{vazquez2008generic,holme2006nonequilibrium,toruniewska2017coupling}, game theory \cite{zimmermann2004coevolution}and
epidemic spreading \cite{gross2006epidemic,marceau2010adaptive,scarpino2016effect,vazquez2016rescue},
to cultural dynamics \cite{sanmiguel2007,centola2007homophily,gracia2011coevolutionary,raducha2017coevolving},
and ecological interactions \cite{jordano2003invariant,guimaraes2011evolution}.

A prototype coevolving model is the Coevolution Voter Model (CVM) combining the voter model and the change of
a network by rewiring links  \cite{vazquez2008generic,holme2006nonequilibrium}. In the original voter model \cite{clifford1973model,holley1975ergodic,suchecki2005voter},
a node can be in one of two states, called up and down. At each step, each node
adopts the state of one of its neighbors chosen randomly.
Adding link rewiring to the original voter model, a coevolving voter model
was addressed \cite{vazquez2008generic,holme2006nonequilibrium}.
In a coevolving voter model, in addition to node's dynamics
following the voter model, the links in a network can be updated.
In a typical CVM, a global rewiring of links is adopted for the sake of
simplicity, meaning that a newly connected node is chosen randomly out of
all nodes in a whole network. The coevolving voter model shows a generic
absorbing phase transition between a connected and fragmented network \cite{vazquez2008generic}.
However, the global rewiring might be unrealistic since there should be
some limitation of the range of searching for a node to be connected.
Many observations indicate that a link rewiring in networks is
mainly implemented by the process of triadic closure, i.e. the tendency that
nodes search for new contacts through existing
neighbors \cite{newman2003social,lee2010,klimek2013triadic}.
In this regard, the effect of triadic closure was analyzed in many models of social
dynamics \cite{bianconi2014triadic,holme2002growing,
davidsen2002emergence,klimek2013triadic}. CVM with rewiring by triadic
closure was also studied \cite{klimek2016dynamical} to describe the community structure of an online society. Triadic closure gives rise to a shattering phase \cite{diakonova2014absorbing}where many lonely nodes separate
from a large connected component.

Recently, in addition to triadic closure another variant of CVM was proposed,
incorporating collective behaviors between connected nodes in terms of
nonlinearity of interactions \cite{min2017fragmentation}.
The nonlinearity means that individuals can take into account the state of
all of their neighbors as a whole, instead of a dyadic interaction, so that
their action does not have to be proportional to the aggregated state
of their neighbors. Modeling of this type of behavior goes generally under the name of Nonlinear Voter Model~\cite{lambiotte2008dynamics,schweitzer2009nonlinear,castellano2009nonlinear,
nyczka2012phase,jkedrzejewski2017pair,radosz2017q,peralta2018analytical}.
In case of nonlinearity, diverse phases depending on the nonlinearity
were found with the different mechanisms of network
fragmentation \cite{min2017fragmentation}.
A similar form of nonlinearity was considered in social impact theory \cite{nowak1990private},
in language competition
dynamics, called volatility \cite{abrams2003linguistics,vazquez2010agent},
or in language evolution problems \cite{nettle1999using}.
However, the effect of nonlinearity on coevolutionary dynamics
has been explored only for a global link rewiring as the simplest example.

In this work, we study a coevolving nonlinear voter model \cite{min2017fragmentation} 
with a local rewiring making triadic
closure \cite{klimek2016dynamical}. These two factors, nonlinearity and triadic 
closure, had been previously analyzed
changing the output of the standard coevolving voter model.
But, the two effects have not been,so far, taken into account together.
We focus on a coevolving nonlinear voter model with a local rewiring
where individuals interact with their neighbors in a nonlinear manner and
a new neighbor can be chosen from among nodes distant by two edges, i.e.
neighbors of neighbors.
We numerically find three possible phases with different topological properties
and magnetization: absorbing consensus phase with a single component, absorbing fragmented
phase with two components of opposite states, and a dynamically active shattered
phase with a coexistence of both states within the main
component and a significant part of isolated nodes. 

\section{Model}\label{section:model}

In our coevolving nonlinear voter model with triadic closure, coevolution is
characterized by a plasticity parameter $p$ defining the ratio between the
timescale of changes in the network topology and the timescale of the dynamics
of nodes' states. A nonlinearity parameter $q$ measures the nonlinear effect
of local majorities in the imitation mechanism of the voter model.
Local rewiring means that when a link from an active node is suppressed, a new link from
this node can only be created with neighbors of its neighbors, without creating
multiple- or auto-connections.

\begin{figure}
\includegraphics[width=\linewidth]{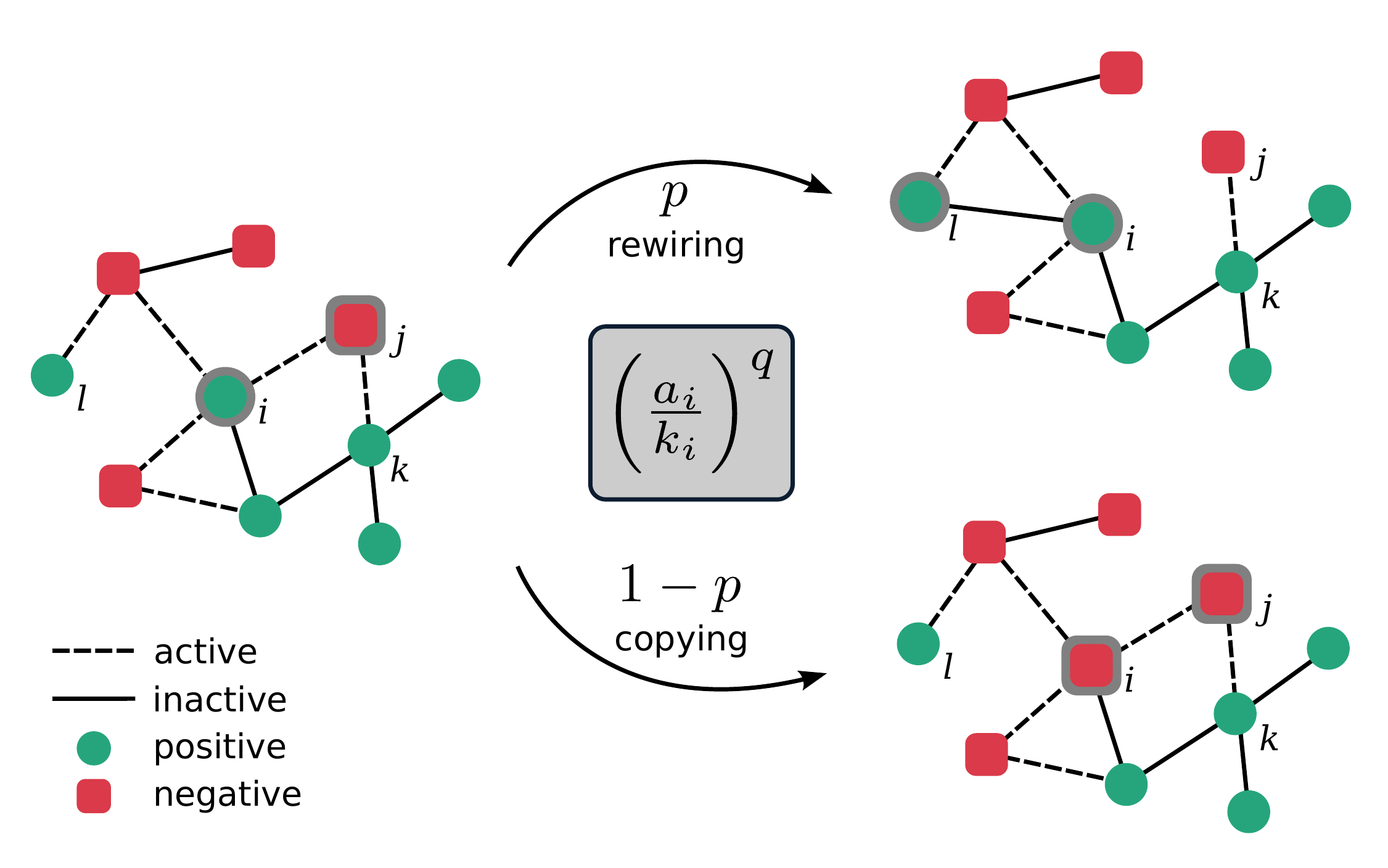}
\caption{Schematic illustration of update rules in a nonlinear coevolving voter
model with triadic closure. Every node is in a state $+1$ (green) or $-1$ (red).
The focal node $i$ is chosen randomly. Then with probability $(\frac{a_i}{k_i})^q$
one of the active links to the node $j$ is chosen. 
With probability $p$ the link ($i$-$j$) is rewired to link ($i$-$l$) (as in the figure)
or to link ($i$-$k$). With probability $1-p$ the focal node $i$ copies the state of node $j$.
}
\label{fig:model}
\end{figure}

Our detailed algorithm is as follows. We start every simulation with
Erd\"os-R\'enyi (ER) graphs \cite{ER1960evolution} with a mean degree $\langle k \rangle=8$.
Every node is initially in one of two possible states $s_i = \pm 1$ with an equal
probability $1/2$. The number of active links $a_i$ of a node $i$ is defined
as the number of connections to nodes in a different state $s_j \neq s_i$.
Consequently, we define the density of active links $\rho_i$ as the fraction of active
links $\frac{a_i}{k_i}$, where $k_i$ is the degree of node $i$.
In every time step a node $i$ is randomly selected from the network (Fig.~\ref{fig:model}).
With probability $\rho_i^q$, where $q$ is a nonlinearity parameter,
the focal node $i$ interacts with its neighbors.
If node $i$ interacts, one of the active links to a neighbor $j$ is chosen at random.
Next, with probability $p$ the active link is rewired, and with probability $1-p$ the
focal node changes its state to the same state of node $j$. In the rewiring process the
chosen active link is suppressed and a new one is created linking $i$ to a node distant
by two links (but not less) and being in the same state as node $i$. If no such
node exists, nothing happens. Obviously, the number of nodes $N$ and number of
links $M$ is constant in time. This triadic closure process is repeated until
a dynamically active stationary state is obtained or an absorbing configuration is reached.

If the plasticity parameter $p=0$ there is no rewiring. Therefore, the network
configuration stays constant throughout the simulation, and fragmentation or
shattering are not possible. On the other hand, for $p=1$ nodes do not change
states so that  consensus cannot be obtained, and links are rewired until the
network undergoes a fragmentation transition into two separate components with
opposite states (with possible detached nodes). The nonlinearity parameter $q$ defines
the likelihood of interacting with individuals in the opposite state. For $q=1$
the model becomes the original coevolving voter model of random imitation of a neighbor.
When $q>1$, nodes with more active links have a higher chance to change their
state than in the ordinary voter model where this frequency is proportional to
the number of active links. In contrast, if $q<1$, nodes with less active links
are more likely to change their state than in the ordinary voter model.

\section{Phase diagram: consensus, fragmentation, and shattering phases}

\begin{figure*}[t]
\centering
  \includegraphics[width=0.32\linewidth]{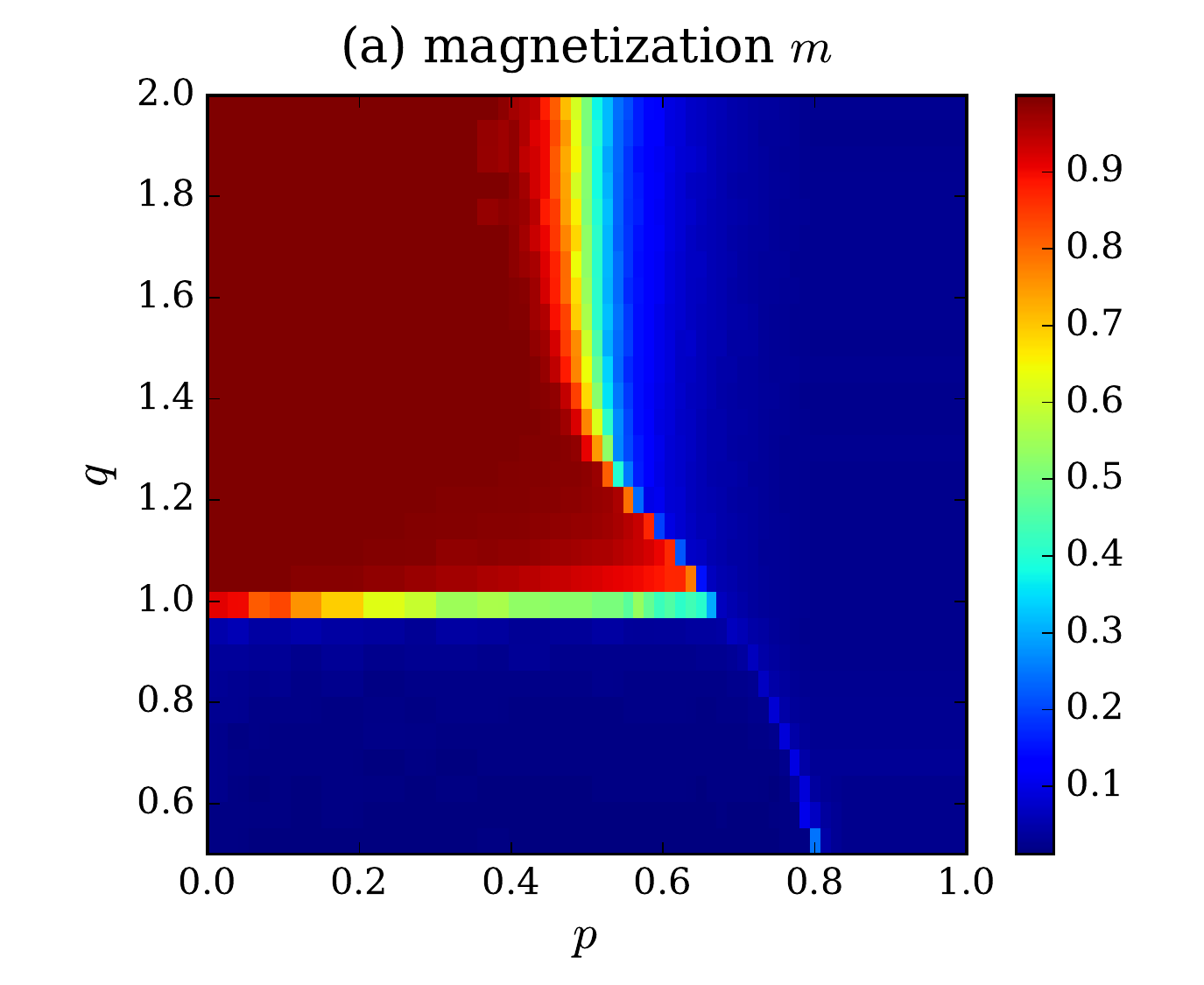}
  \includegraphics[width=0.32\linewidth]{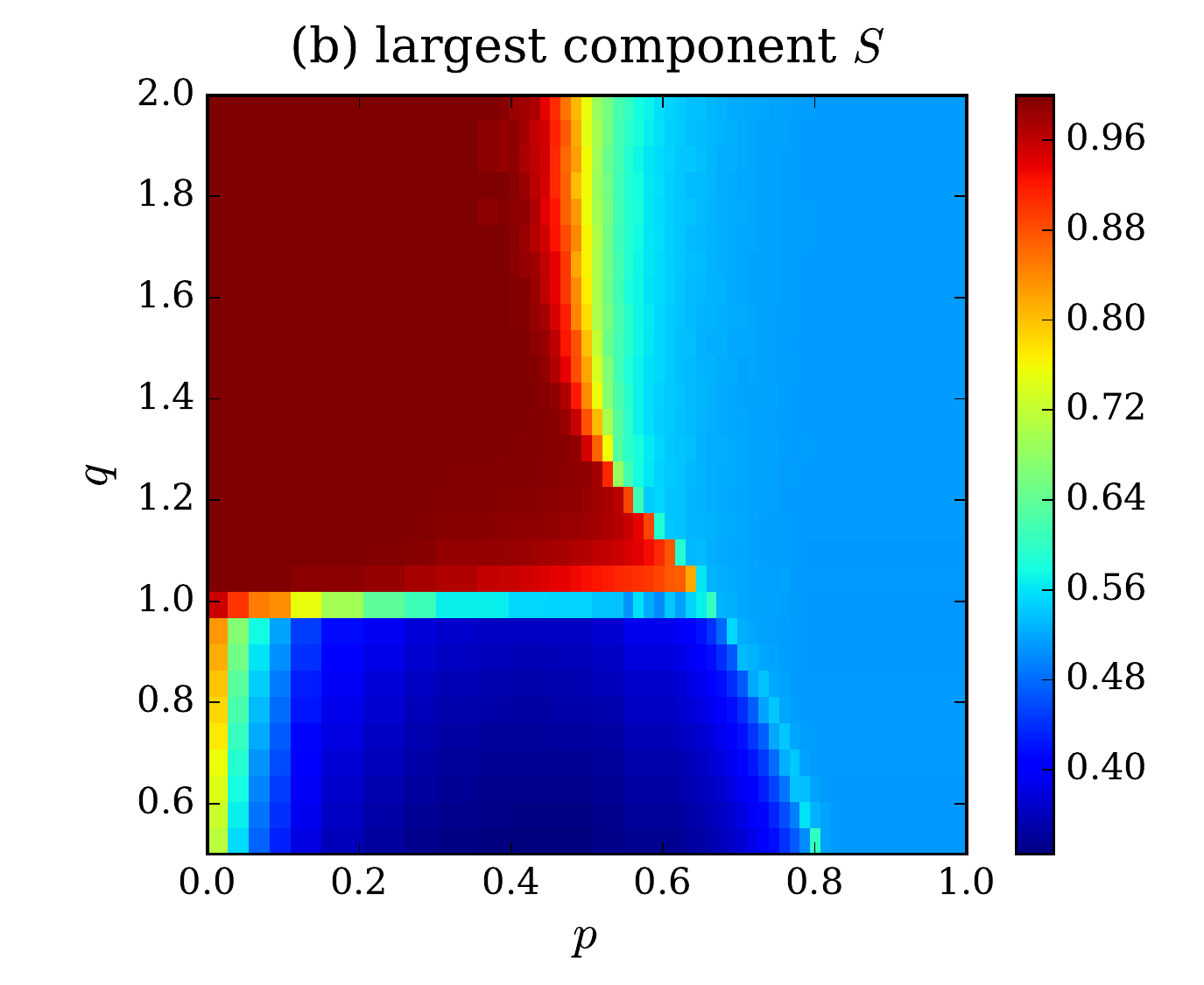}
  \includegraphics[width=0.32\linewidth]{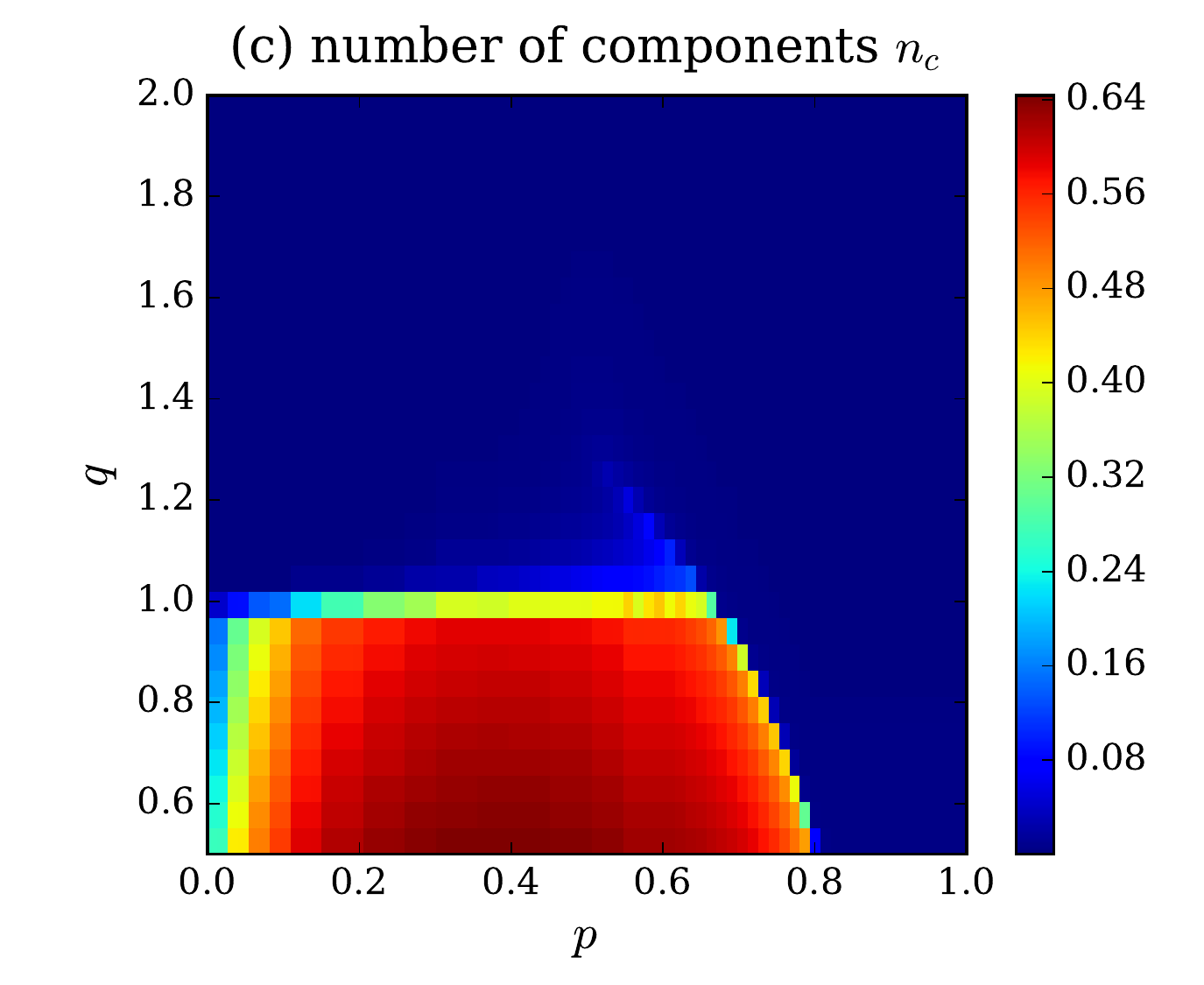}
\caption{Phase diagram in $(p,q)$ space.
(a) Magnetization, (b) the size of the largest component, and (c) the number of components on ER networks
with $N=5000$, $M=20000$, averaged over $500$ simulation runs.
}
\label{fig:phase}
\end{figure*}

\begin{figure}
\centering
 \includegraphics[width=1.0\linewidth]{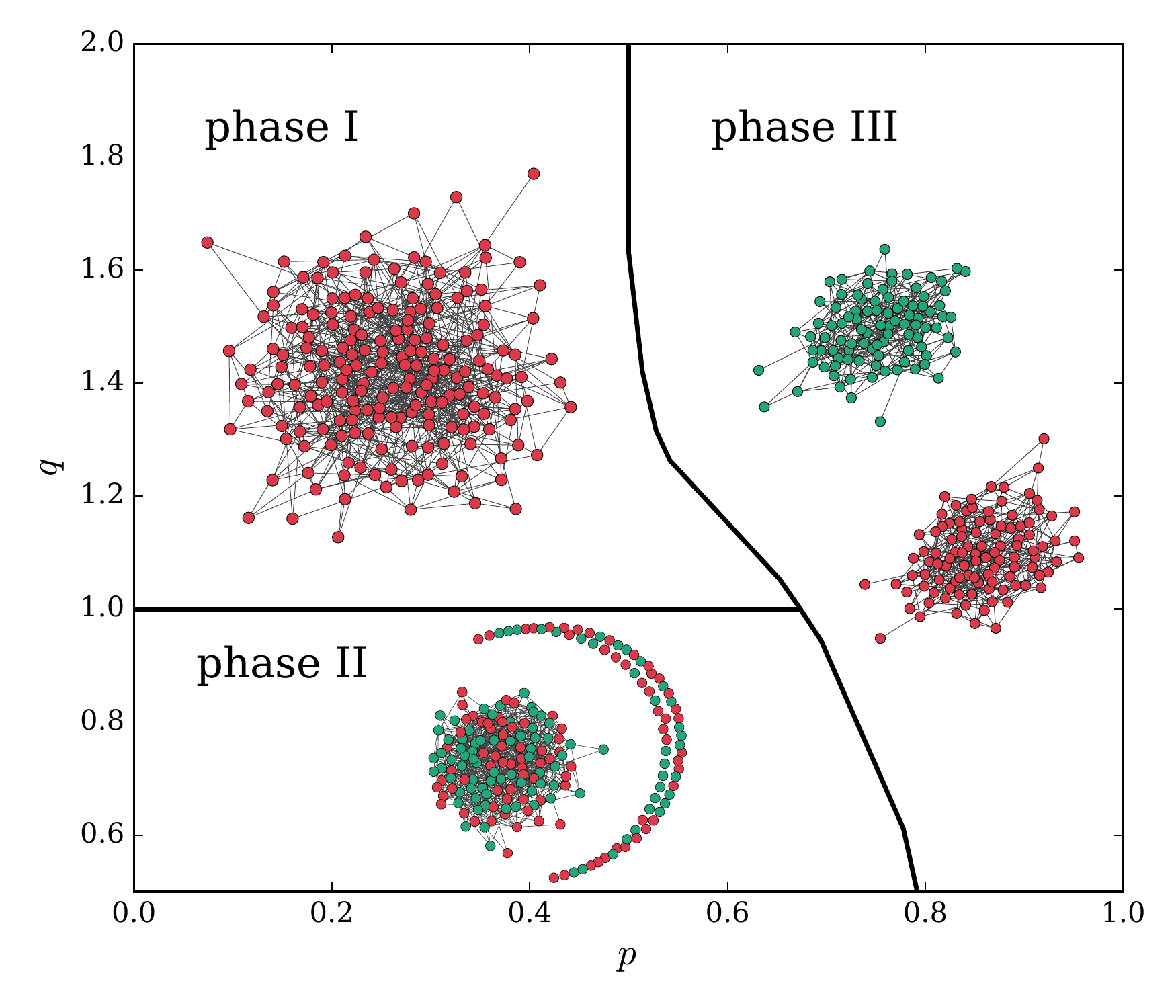}
\caption{Scheme of the phase diagram with respect to $p$ and $q$.
Different phases are separated by a black line and for each a typical final
configuration is presented, with opposite states colored in red and green.
}
\label{fig:network}
\end{figure}

Our simulations are described in terms of three quantities: the magnetization
$m = \frac{1}{N} \sum_i s_i$, the relative size of the largest connected component $S$ of
the network, and the relative  number of  separate components $n_c$. Phase diagrams for these
quantities in the $(p,q)$ parameter space are shown in Fig.~\ref{fig:phase}. We find three
different steady-state phases.
For $q>1$ and $p<p_c(q)$, where $p_c(q)$ is the transition line separating the phase III from others, phase I -- an absorbing 
consensus phase -- is characterized
by a non-zero value of the absolute magnetization and a single large component,
$(|m|,S,n_c)\approx (1,1,0)$. When $q<1$ and $p<p_c(q)$, we obtain phase II, a shattered phase, where
the absolute magnetization drops to zero and the network is composed of an active
component and a number of shattered nodes.
In the shattered phase, many nodes that initially belong to a
connected component become shattered into isolated nodes,
resulting in a high value of $n_c$.
Therefore, the phase II can be identified as
$(|m|,S,n_c)\approx (0,S^*,n_c^*)$ where $S^*$ is the size of a large active component of the network
and $n_c^*$ is the number of shattered nodes.
In our parameter sets, $S^*$ and $n_c^*$ are respectively approximately $0.4$ and $0.6$.
Increasing the value of $p$ above $p_c$ ($p>p_c$) for any $q$, we obtain
phase III, a fragmented phase characterized by $(|m|,S,n_c)\approx (0,1/2,0)$.
In phase III, the network fragments in two components of approximately the
same size and each component in an opposite consensus state, so that the absolute magnetization
is close to zero and the size of the largest component is around 1/2.

The topological properties of the three different phases are illustrated
in an example of a network configuration in the steady state (Fig.~\ref{fig:network}).
First, in the consensus phase (phase I), a network is formed by a single connected
component within a full consensus either $m=1$ or $m=-1$ in the steady state.
Second, in the shattered phase (phase II), we find even more than a half of the nodes being
isolated. In addition, the main cluster remains active,
mixing up and down states of nodes at the steady state.
This shattered phase is different from the active state with a connected component
observed in the nonlinear voter model with global rewiring with similar
parameters $(p,q)$ \cite{min2017fragmentation}.
Last, in the fragmented phase (phase III) the network is polarized: two internally coherent clusters in opposite states.

\begin{figure*}[t]
\centering
\includegraphics[width=0.32\linewidth]{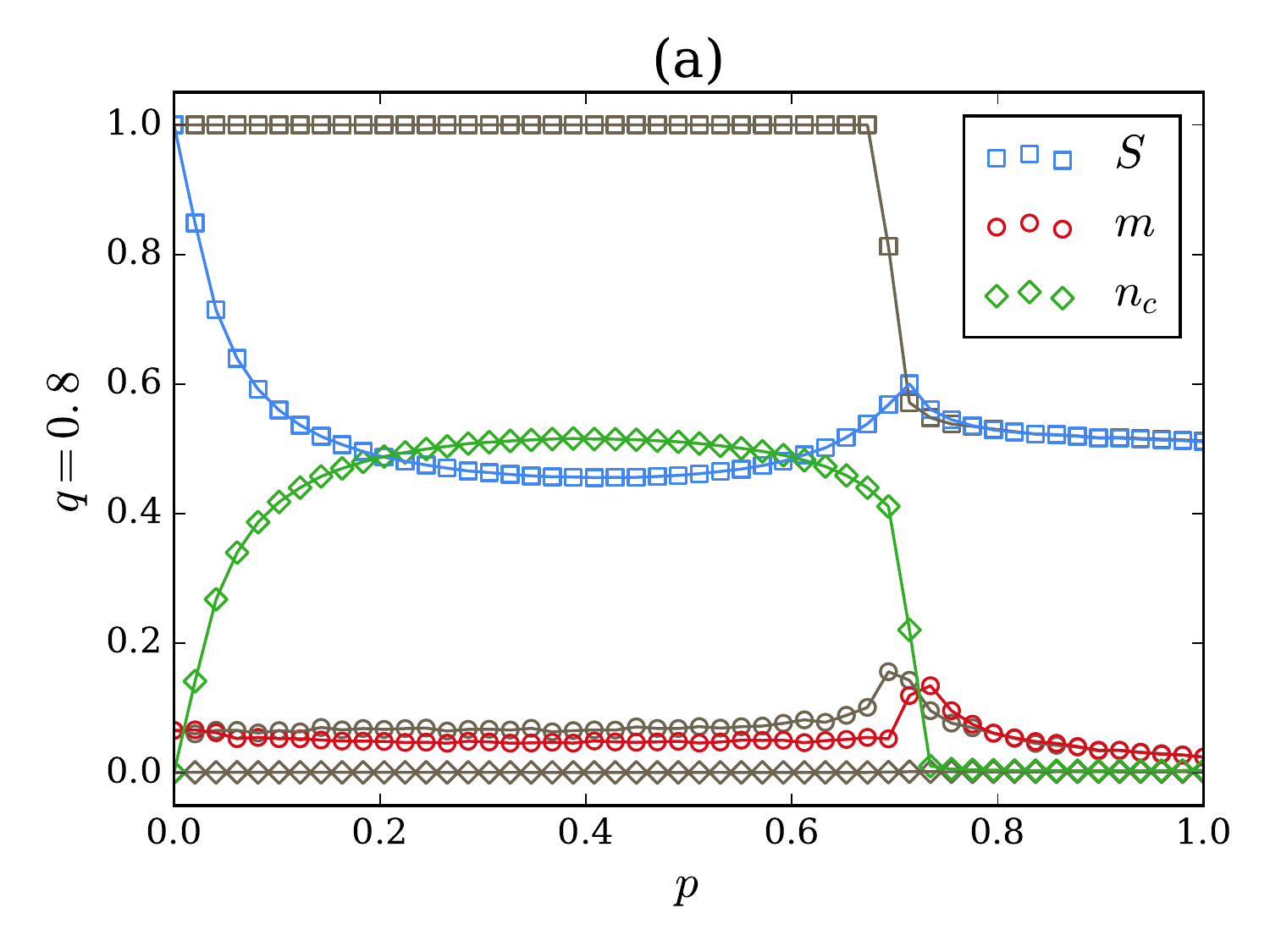}
\includegraphics[width=0.32\linewidth]{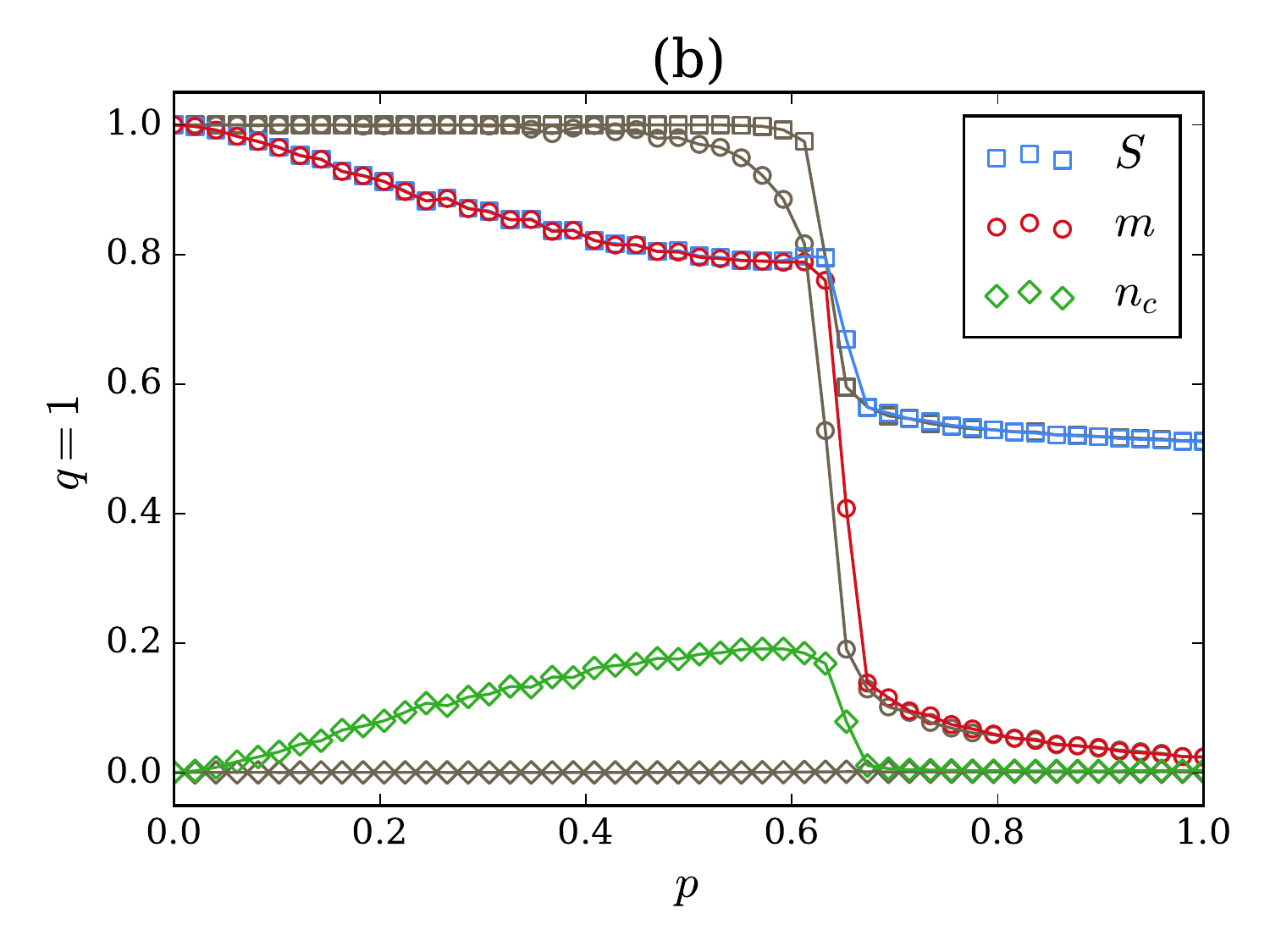}
\includegraphics[width=0.32\linewidth]{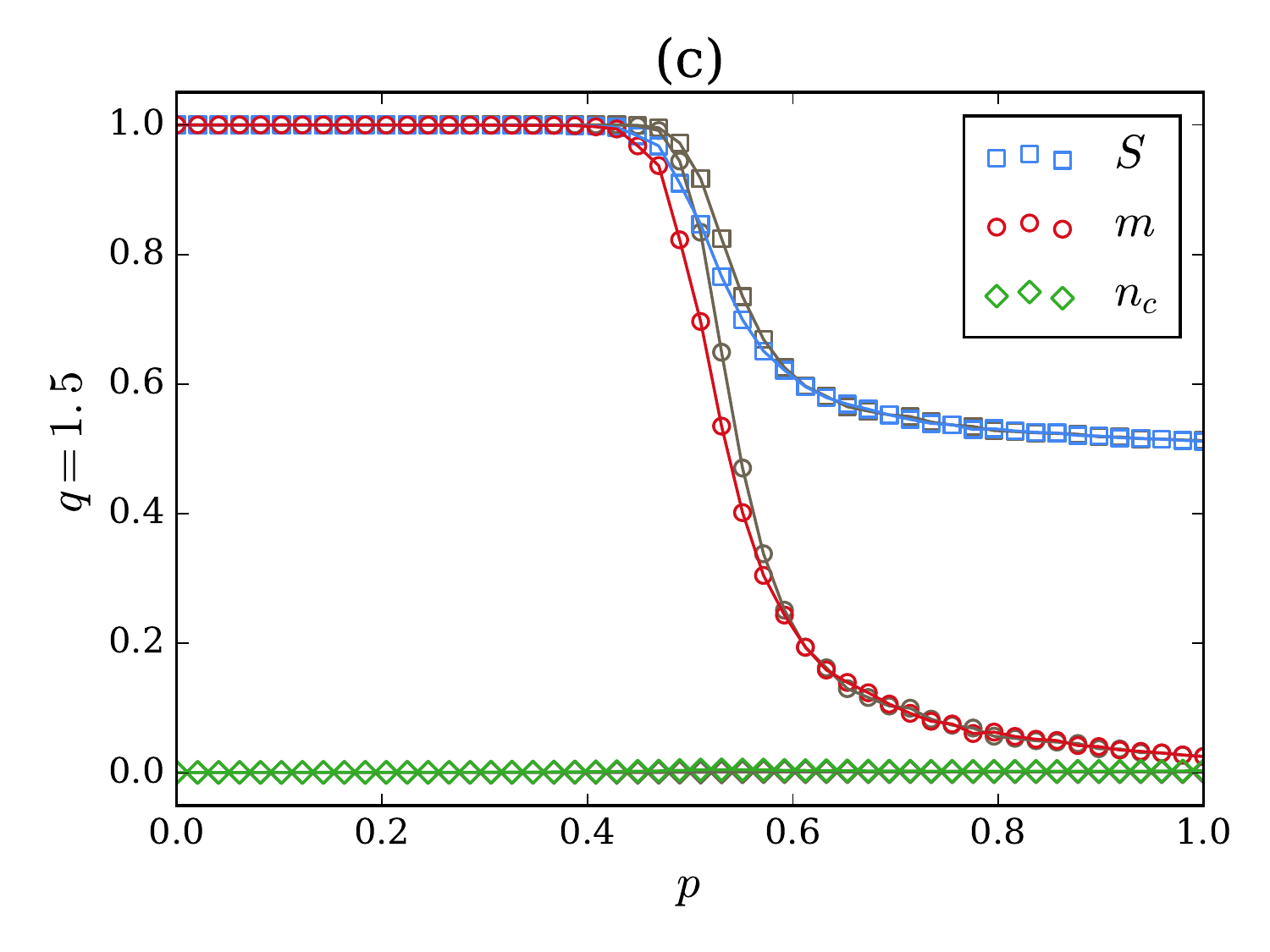}
\includegraphics[width=0.32\linewidth]{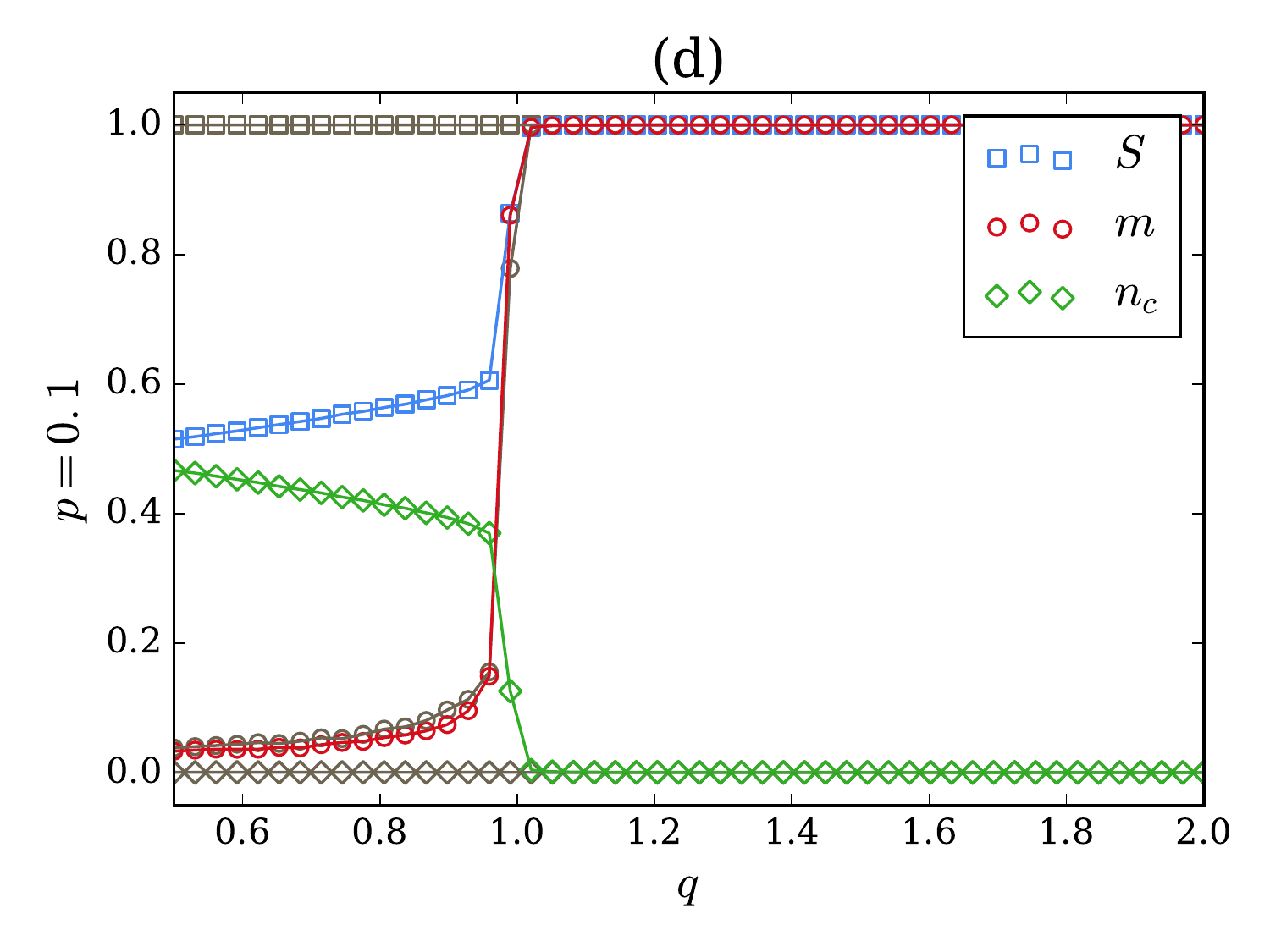}
\includegraphics[width=0.32\linewidth]{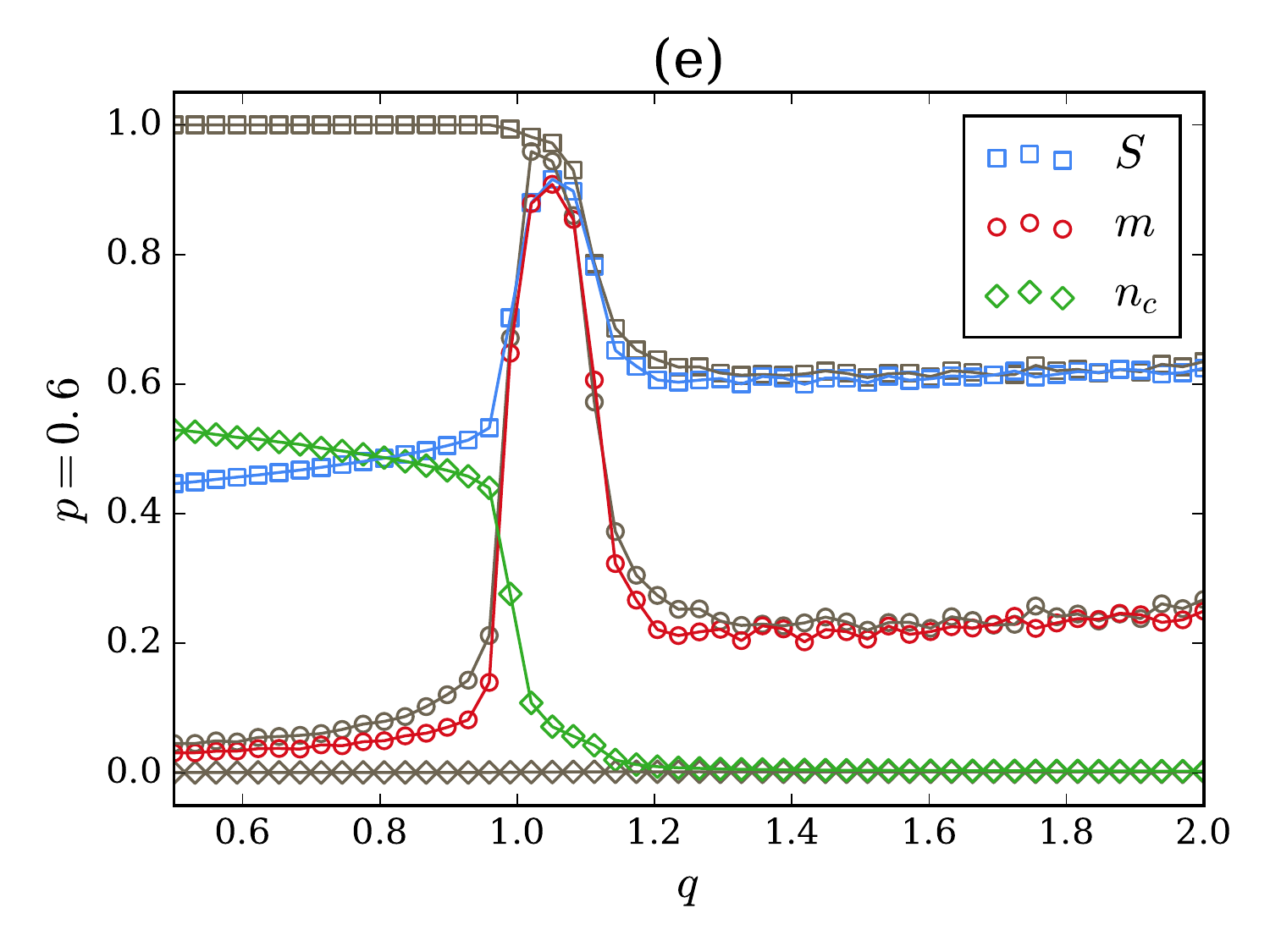}
\includegraphics[width=0.32\linewidth]{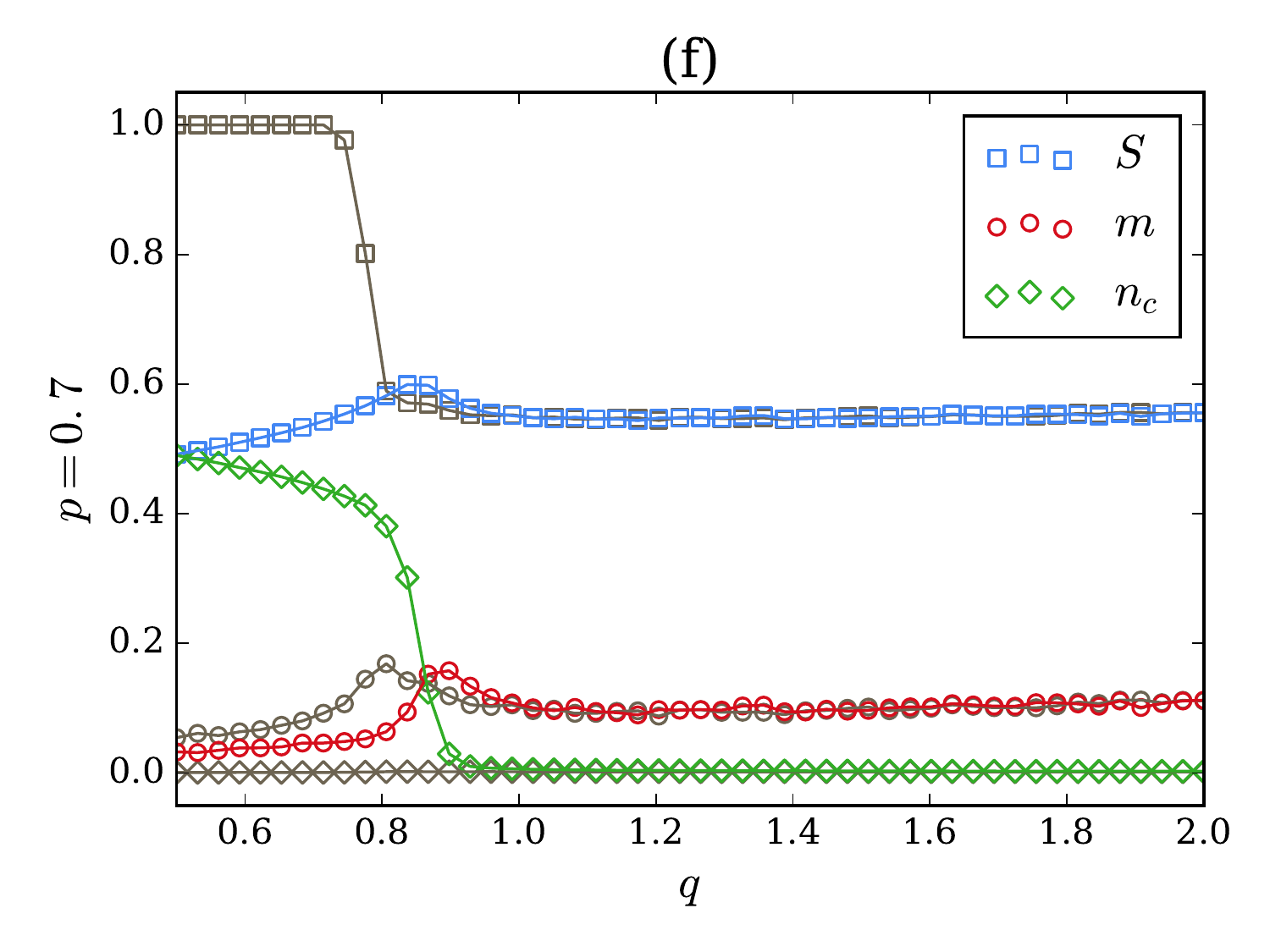}
\caption{Phase transition with respect to $p$ and
	fixed values of $q$ (a) $q=0.8$, (b) $q=1$, and (c) $q=1.5$, and
	with respect to $q$ and fixed values of $p$ (d) $p=0.1$, (e) $p=0.6$, and (f) $p=0.7$.
	The size of the largest component $S$ (blue squares), the magnetization
	$m$ (red circles), and the number of components $n_c$ (green diamonds) on
	ER networks with $N=1000$, $M=4000$, averaged over $500$ simulation runs.
	Every quantity for global rewiring \cite{min2017fragmentation} is also plotted
	with the same symbols in gray.
}
\label{fig:transition}
\end{figure*}

\section{Fragmentation and shattering phase transitions}

In the coevolving nonlinear model with triadic closure, phase transitions between
different phases can be identified by examining $(|m|,S,n_c)$
as a function of $p$ and $q$ (Fig.~\ref{fig:transition}).
When $q<1$ with varying $p$, we find a phase transitions between the shattering (phase I)
and fragmentation (phase III) phases [Fig.~\ref{fig:transition}(a)].
At the critical point $p_c$, a shattered network
with many detached nodes starts to recombine into two clusters with increasing $p$.
In other words, a network with many isolated nodes and an active component
transforms into two connected components with polarized configuration.
The transition can be identified by the sharp drop of $n_c$ with increasing $p$.
Note again that the shattered phase is different from the coexistence phase observed
in an ordinary coevolving nonlinear voter model with global rewiring the network \cite{min2017fragmentation} when $q<1$ and $p<p_c$, as shown
in the differences in $S$ and $n_c$.

When $q>1$, we find a fragmentation transition between the consensus (phase I) and
fragmented (phase III) phases, also observed in the ordinary CVM and
nonlinear CVM [Fig.~\ref{fig:transition}(c)]. Increasing $p$, the network
splits from a single component with a full consensus state into two separate clusters in opposite consensus states. For $p>p_c$, the magnetization
$|m|$ and the number of components $n_c$ approach to zero, and the size of
the largest component $S$ tends to $0.5$. The fragmented phase
(phase III) can be obtained for any value of $q$ if the plasticity
parameter $p$ is higher than $p_c(q)$.
Thus, when $p>p_c$, the behaviors of $S$, $|m|$, and $n_c$
for different $q$ share common features.

In Fig.~\ref{fig:transition}(d-f), we show $S$, $|m|$, and $n_c$ as a function of 
$q$ for fixed values of  $p$ ($p=0.1,0.6,0.7$).
For $p=0.1$ as a representative example of $p<p_c$ [Fig.~\ref{fig:transition}(d)], by
varying $q$ we observe
an absorbing phase transition between phase I and phase II.
For $q \geq 1$, numerical simulations always reach a dynamically frozen state.
For $q < 1$, however we find a dynamically active phase with a high fraction of
isolated nodes and an active cluster showing coexistence of up and down states
of nodes. This shattered phase observed for local rewiring is a new phenomenon
that was not detected for global rewiring.

For $p=0.6$ [Fig.~\ref{fig:transition}(e)], varying $q$
we can observe all three phases and two
transitions. First, for small $q$ we observe the active shattered phase. 
Increasing $q$ above $1$ we first obtain a transition to the consensus phase
with high magnetization $|m|$ and a high value of the largest component size $S$. 
The second fragmentation transition occurs for larger values of $q$ to the fragmented phase
with low values of  $|m|$ and $S\approx 1/2$.
When $p$ increases further, i.e, $p=0.7$ [Fig.~\ref{fig:transition}(f)],
the consensus phase disappears and the shattered phase changes
directly to the fragmented phase as $q$ increases, as indicated
by a decreasing value of $n_c$ at the transition.


\begin{figure}
\centering
\includegraphics[width=0.49\linewidth]{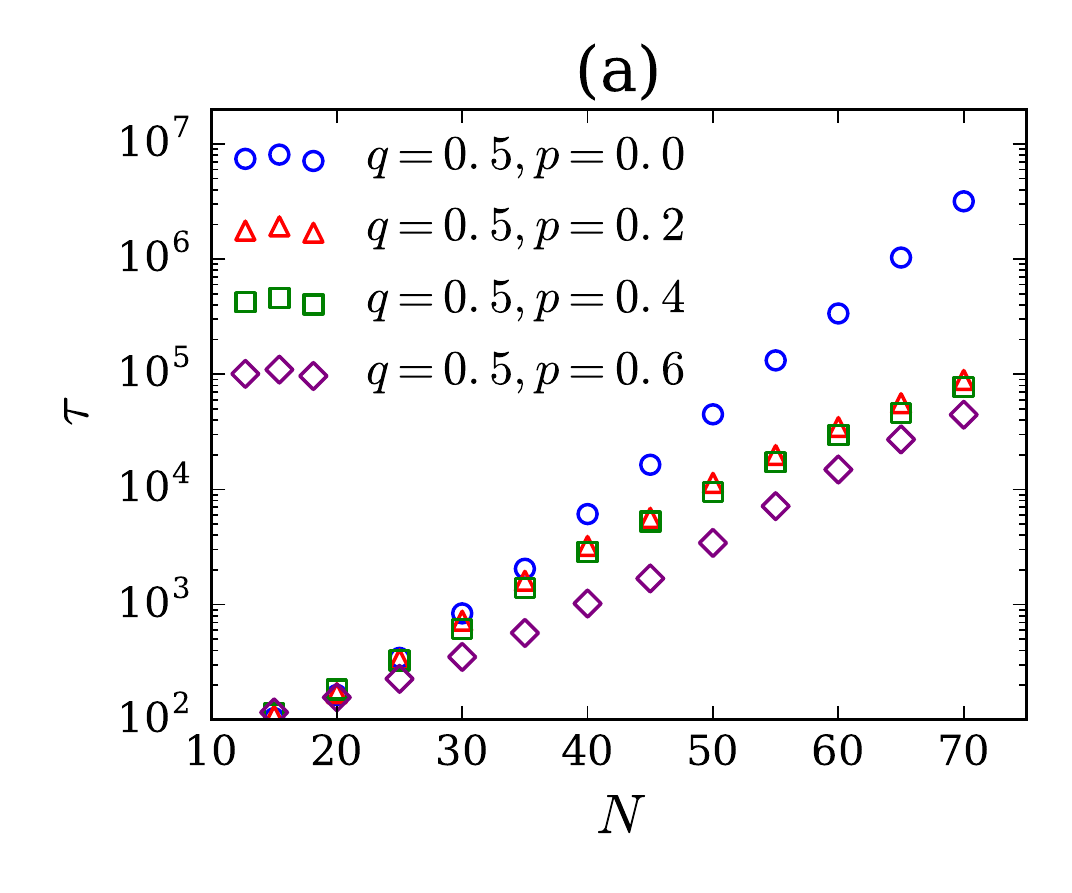}
\includegraphics[width=0.49\linewidth]{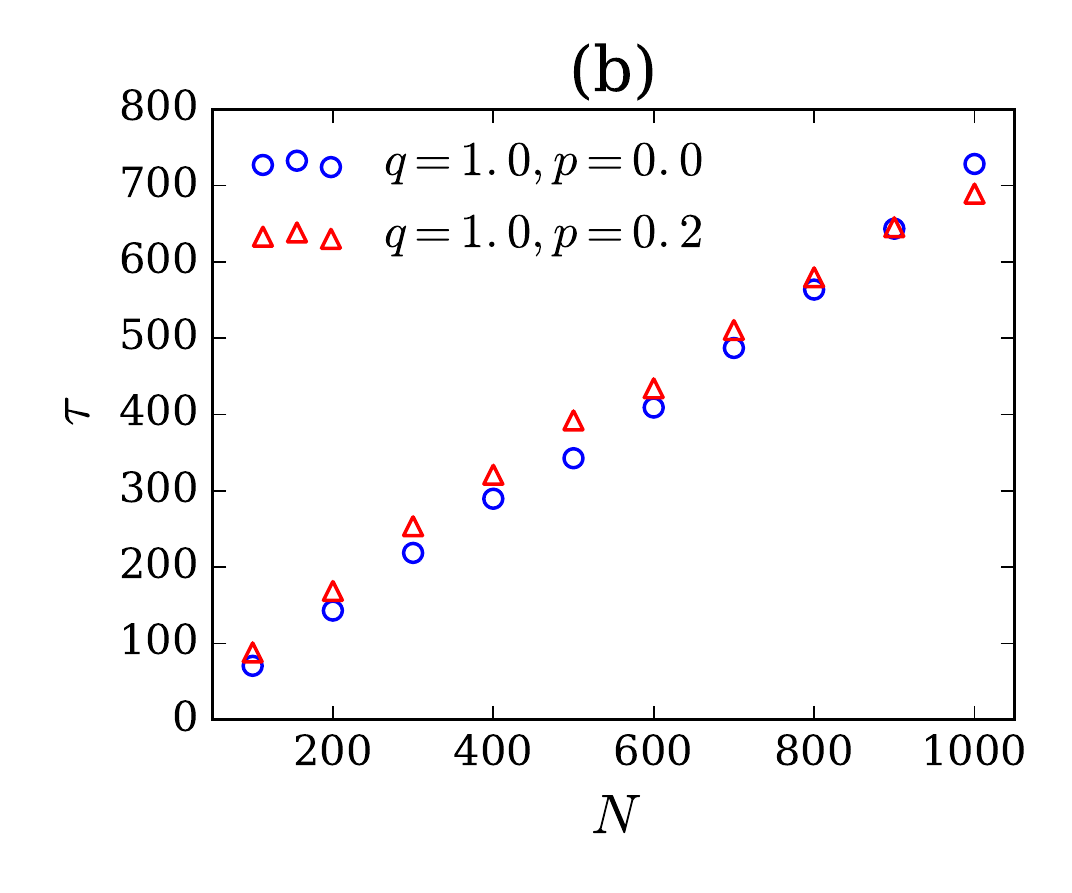}
\includegraphics[width=0.83\linewidth]{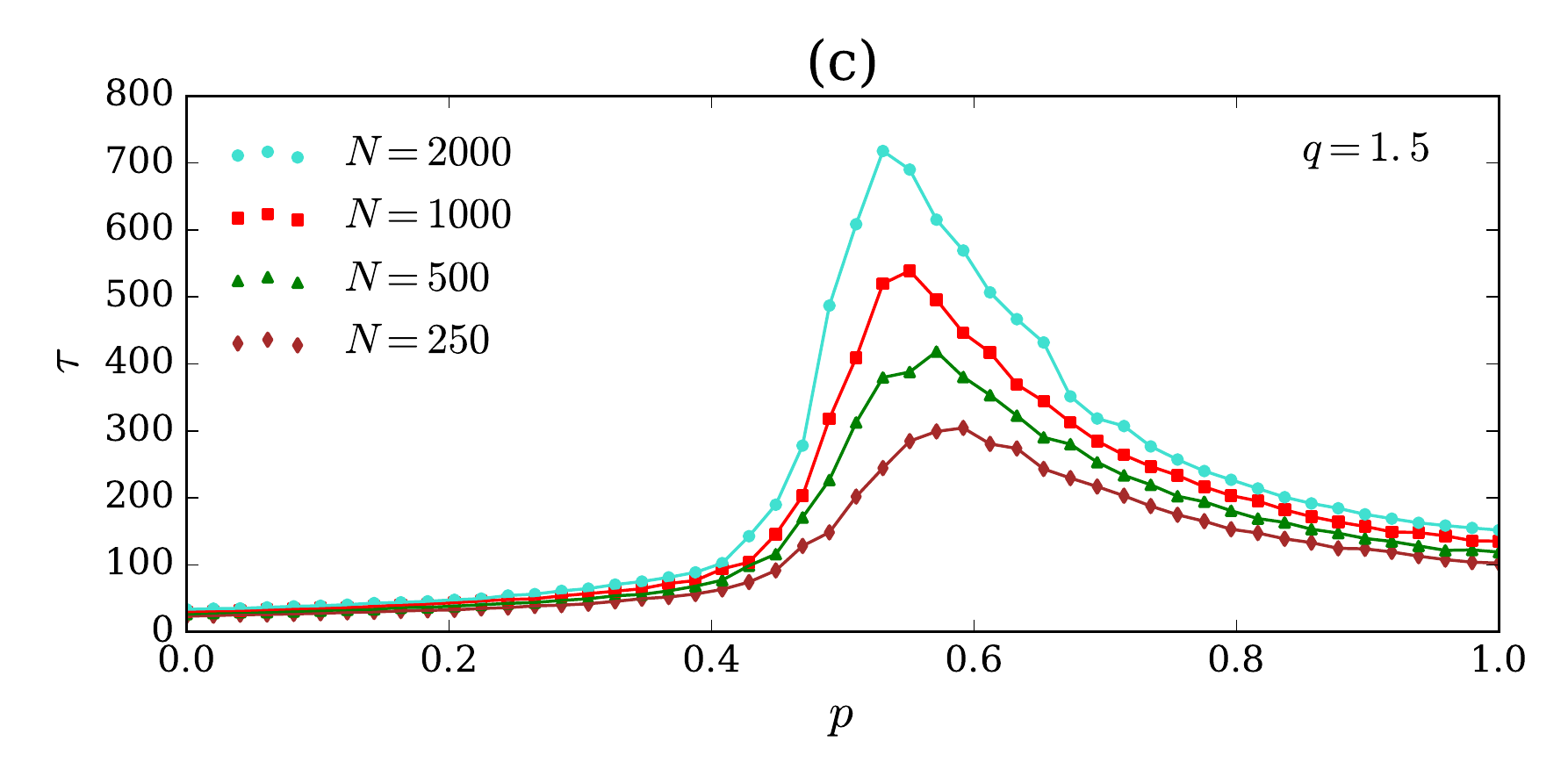}
\caption{Convergence time $\tau$ in the number of Monte Carlo steps to a frozen
	configuration as a function of network size $N$ in (a) the phase II and
	at (b) $q=1$. (c) Convergence time $\tau$ as a function of $p$ for $q=1.5$ 
	(phases I and III). Results are obtained for ER networks with $\langle k \rangle = 8$,
	averaged over 500 realizations.
}
\label{fig:time}
\end{figure}

\section{Convergence time}

A finite system is always bound to reach an absorbing phase for
any combination of parameters in the limit $t \rightarrow \infty$.
But for the phase II, which is an active phase, the convergence time $\tau$ to
an absorbing state grows exponentially with increasing system size $N$,
$\tau \sim e^N$ as in the case of global rewiring \cite{min2017fragmentation}.
Exponential divergence of $\tau$ in Monte-Carlo steps for phase II is
clearly shown in Fig.~\ref{fig:time}(a).
Therefore, it is expected to remain the active phase rather than to reach
an absorbing state in a finite time in the limit $N\rightarrow \infty$
when $q<1$ and $p<p_c$.
For $q=1$ (corresponding to a linear interaction) and $p<p_c$, we reproduce a
linear scaling with system size $N$, $\tau \sim N$, see Fig.~\ref{fig:time}(b),
as reported also in the linear CVM and nonlinear CVM with global rewiring
\cite{vazquez2008generic,min2017fragmentation}.
We also find that the convergence to the frozen state with $q=1.5$ slows down 
at the critical point $p_c$ [Fig.~\ref{fig:time}(c)].




\begin{figure}
\centering
\includegraphics[width=0.49\linewidth]{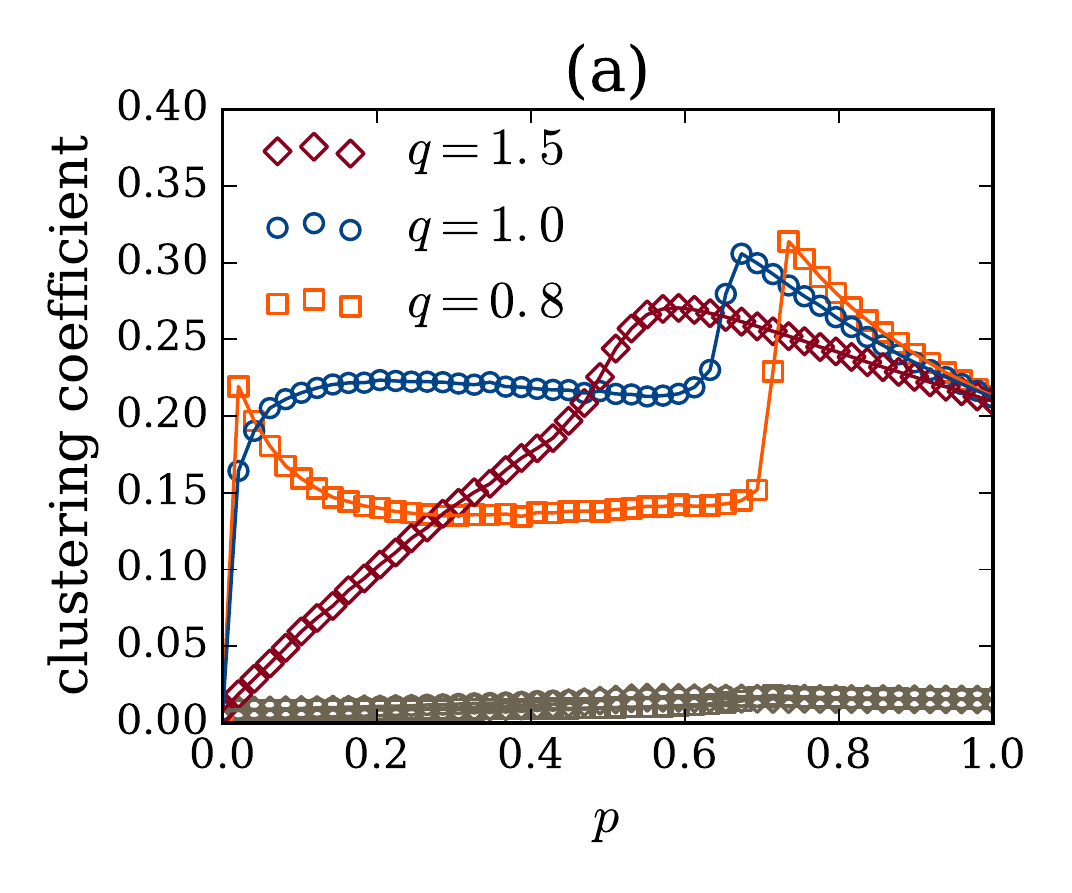}
\includegraphics[width=0.49\linewidth]{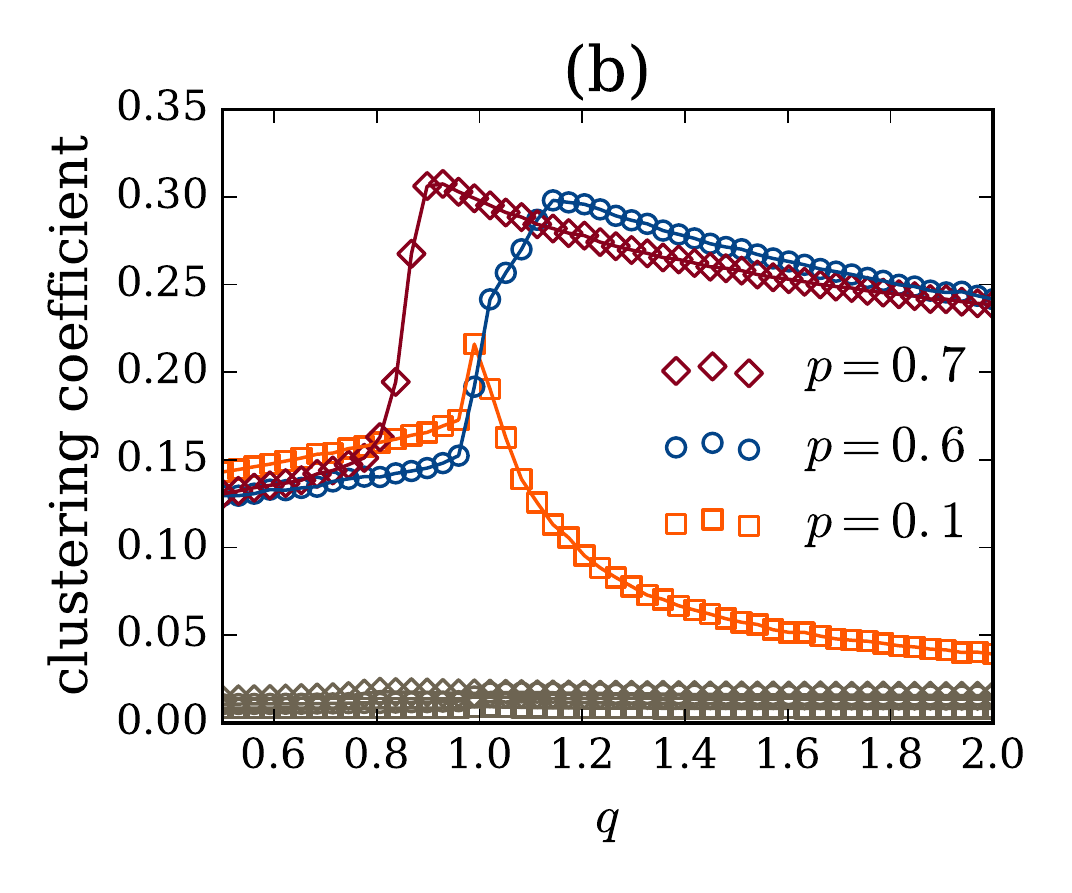}
\caption{Average local clustering coefficient $\langle C \rangle$ as a function of (a) $p$ and (b) $q$,
	on ER networks with $N=1000$, $M=4000$, averaged over $500$ simulation runs.
	Every quantity for global rewiring \cite{min2017fragmentation} is also plotted
	with the same symbols in gray.
}
\label{fig:cluster}
\end{figure}

\section{Clustering coefficient}

We finally investigated the internal structure of networks generated in the model
focusing on the average clustering coefficient $\langle C \rangle$. 
For a node $i$ with degree $k_i$, the local clustering coefficient $C_i$
is defined as \cite{watts1998collective}
\begin{align}
C_i = \frac{2 L_i}{k_i (k_i-1)}
\end{align}
where $L_i$ is the number of links between the $k_i$ neighbors of node $i$.
Then, the average clustering coefficeint representing the average 
of $C_i$ over all nodes is given by
\begin{align}
\langle C \rangle &=\frac{1}{N} \sum_{i=1}^N C_i.
\end{align}
In Fig.~\ref{fig:cluster} we present average
local clustering coefficient for different values of $(p,q)$ in the steady state.
For nodes having less than two links, local clustering coefficient is not well defined,
and for that reason we exclude these nodes in our analysis.
While the value of the clustering coefficient for global rewiring remains almost zero
for all tested parameter sets, it clearly shows values far from zero, exceeding $0.3$
in an extreme case, for local rewiring.
In addition, the location of peaks in $\langle C \rangle$ is coincident with
the transition point, meaning that the clustering coefficient can be an
indicator of phase transitions.
Considering the fact that real-world social systems are known to display
high clustering \cite{albert2002statistical,newman2003social,klimek2013triadic},
our model can provide a plausible way to reconstruct such structure
aiming to describe social phenomena.

\section{Discussion}\label{section:methods}

In this paper, we have studied a coevolving nonlinear voter model with local 
rewiring. We identify three different phases, namely consensus, fragmented
and dynamically active shattered phases, characterized by different topological structures and
their global magnetization. We also examine the transitions between these
phases in terms of the size of the largest cluster, the number of 
clusters, and the magnetization. We find an active shattered
phase with $q<1$ and $p<p_c$ where the majority of nodes are isolated while
a large cluster remains active. This distinct phase which is not detected
in global rewiring implies that local rewiring might be the origin
of many isolated parts in complex adaptive systems, such as social systems.
It would be also interesting
\cite{diakonova2015noise,carro2016noisy,peralta2018analytical}
to investigate the
influence of noise that is a source of randomness, and multilayer
structures with the nonlinearity and/or triadic closure
\cite{diakonova2014absorbing}.

\acknowledgments
We acknowledge financial support from the Agencia Estatal de Investigaci on
(AEI, Spain) and Fondo Europeo de Desarrollo Regional under project ESOTECOS
FIS2015-63628-C2-2-R (MINECO/AEI/FEDER,UE). 
This work was supported by the National Research Foundation of Korea (NRF)
grant funded by the Korea government (MSIT) (No. 2018R1C1B5044202).

\bibliographystyle{eplbib.bst}

\end{document}